\documentclass[11pt,twoside]{article}

\usepackage{asp2006}
\usepackage{epsf}
\usepackage{psfig}
\usepackage{lscape}

\begin{document}
\title{Hot Subdwarfs in Binaries as the Source of the Far-UV Excess in
Elliptical Galaxies } 

\author{Philipp Podsiadlowski,\altaffilmark{1} Zhanwen Han,\altaffilmark{2}
Anthony E. Lynas-Gray,\altaffilmark{1} and David Brown\altaffilmark{1}}
\altaffiltext{1}{University of Oxford, Department of Physics, Oxford, 
OX1 3RH, UK}
\altaffiltext{2}{National Astronomical Observatories / Yunnan Observatory,
the Chinese Academy of Sciences, P.O.Box 110, Kunming, 650011, China} 
\setcounter{page}{15}


\begin{abstract} 
The excess of far-ultraviolet (far-UV) radiation in elliptical
galaxies has remained one of their most enduring puzzles. In contrast,
the origin of old blue stars in the Milky Way, hot subdwarfs, is now
reasonably well understood: they are hot stars that have lost their
hydrogen envelopes by various binary interactions. Here, we review the
main evolutionary channels that produce hot subdwarfs in the Galaxy
and present the results of binary population synthesis simulations
that reproduce the main properties of the Galactic hot-subdwarf
population. Applying the same model to elliptical galaxies, we show
how this model can explain the main observational properties of the
far-UV excess, including the far-UV spectrum, without the need to
invoke {\em ad hoc} physical processes.  The model implies that the UV
excess is not a sign of age, as has been postulated previously, and
predicts that it should not be strongly dependent on the metallicity
of the population.
\end{abstract}


\section{Introduction}

One of the first major discoveries soon after the advent of UV
astronomy was the discovery of an excess of light in the
far-ultraviolet (far-UV) in elliptical galaxies (see the review by
O'Connell 1999). This came as a complete surprise since elliptical
galaxies were supposed to be entirely composed of old, red stars and
not to contain any young stars that radiate in the UV.  Since then it
has become clear that the far-UV excess (or upturn) is not a sign of
active contemporary star formation, but is mainly caused by an older
population of helium-burning stars or their descendants with a
characteristic surface temperature of 25,000\,K (Ferguson et al.\
1991), also known as hot subdwarfs or sdB stars. In recent years, it
has become increasingly clear that their Galactic counterparts are
predominantly produced by binary interactions (Maxted et al.\ 2001;
Morales-Rueda et al.\ 2003; Lisker et al.\ 2005; Green 2008).  Here,
we first review the main evolutionary channels that produce hot
subdwarfs in our Galaxy (see Han et al.\ 2002, 2003 for further
details) and then apply this model to elliptical galaxies to show that
these may also be able to account for their far-UV excess (Han,
Podsiadlowski, \& Lynas-Gray 2007).

\section{Evolutionary Channels for sdB Stars}

To produce a helium-core-burning sdB star, it is necessary that the
progenitor of the sdB star loses its hydrogen envelope just before
reaching the tip of the first giant branch (FGB). In a {\em
single-star scenario for sdB stars} (e.g., D'Cruz et al.\ 1996), this
requires highly variable mass loss on the FGB, where some stars
experience enhanced mass loss on the FGB (e.g., due to helium mixing
driven by internal rotation [Sweigart 1997]) or at the helium flash
itself; e.g. Han et al.\ (1994) showed that 1-$M_\odot$,
solar-metallicity giants have only marginally bound envelopes near the
tip of the FGB and that it is much easier to eject their envelopes at
the helium flash than for their more massive counterparts with more
tightly bound envelopes.

In a {\em binary scenario}, the progenitor of the sdB star typically has
to fill its Roche lobe near the tip of the FGB to lose most of its
hydrogen-rich envelope in the ensuing binary interaction. This can occur
either through stable Roche-lobe overflow (RLOF) or in a common-envelope
(CE) phase (Paczy\'nski 1976). Alternatively, a hot subdwarf can
be produced by the merger of two helium white dwarfs if helium
is ignited in the merger product.

In order to ignite helium in the core and therefore have a long-lived
subdwarf phase lasting $\sim 10^8\,$yr, one has to distinguish whether
the progenitor develops a degenerate helium core or not. 
Relatively low-mass stars ($M_0\la 2\,M_\odot$ for models with
moderate convective overshooting) develop degenerate helium cores,
and helium will only be ignited if the core mass at the
beginning of the mass-transfer phase is quite close to the core mass at
the helium flash (typically within 0.025$\,M_\odot$) This core mass
range corresponds to a range of red-giant radii $\Delta \log R \simeq
0.07$, which can be related to a range of initial orbital periods at
which a typical giant in a binary fills its Roche lobe of $\Delta \log P
\simeq 0.1$.  For a canonical binary orbital-period distribution, this
implies that $\sim 1\,$\% of all stellar systems in the appropriate
mass range will lose their envelopes at this stage by binary
interactions. This corresponds to a characteristic birthrate for
sdB star in the Galaxy of $10^{-2}\,$yr$^{-1}$, consistent with
previous observational estimates (e.g., Heber 1986).

Stars with an initial mass $M_0\ga 2\,M_\odot$ ignite helium under
non-degenerate conditions, and mass loss can occur at any time after
the main-sequence phase. In contrast to their lower-mass counterparts
which have a sharply peaked sdB mass distribution around 0.46\,$M_\odot$,
these produce a much wider distribution of sdB star masses, which can
be as low as $0.32\,M_\odot$.

\begin{figure}[t]
\hspace{-2.5cm}\psfig{figure=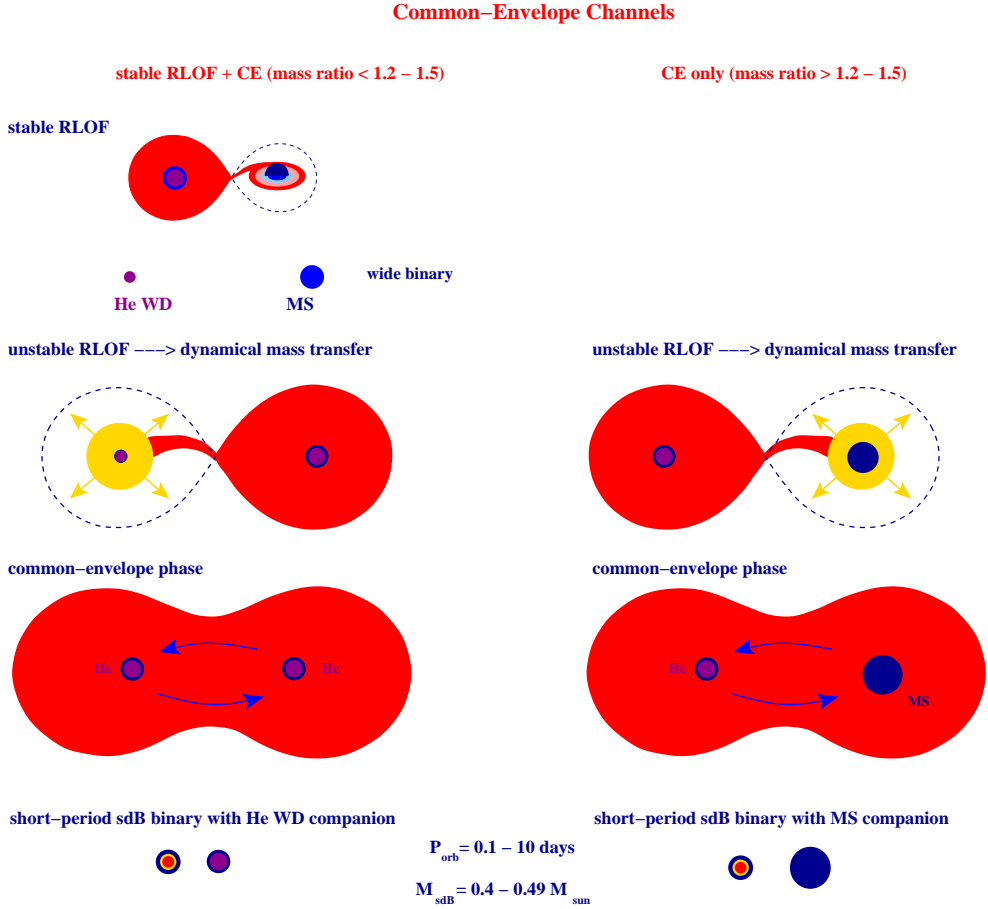,width=1.35\textwidth,angle=-90}
\caption{Common-envelope (CE) channels for the production of hot
subdwarfs.  The CE phase can be either the first {\em (right panel)\/}
or the second {\em (left panel)} mass-transfer phase, producing a
tight sdB binary with either a main-sequence or white-dwarf companion,
respectively.}
\end{figure}

\subsection{Common-Envelope Channels}

In the CE channel (see Figure~1), the progenitor of the sdB star is a
giant that fills its Roche lobe near the tip of the FGB and
experiences dynamical mass transfer. This generally requires that the
mass ratio of the donor star to the secondary, the accreting
component, is larger than $\sim 1.2$.  In this case, the secondary is
unable to accrete all of the transferred matter and starts to fill and
ultimately overfill its own Roche lobe. This leads to the formation of
a common envelope surrounding both stars. Inside the common envelope,
the core of the giant and the secondary form an immersed
binary. Because of friction with the envelope, these two components
spiral towards each other until enough orbital energy is released to
eject the envelope (Paczy\'nski 1976). This ends the spiral-in phase
and leaves a much closer binary with an orbital period typically
between 0.1 and 10\,d, consisting of the core of the giant and the
secondary.

In general, the CE phase can be either the first or the second
mass-transfer phase in binary. In the former case, the companion of
the sdB star is expected to be a normal star, most likely a
main-sequence star, while in the latter case it is a helium white
dwarf.  In the best-fit model of Han et al.\ (2003), the birthrates of
sdB stars in the CE channel are $\sim 0.7\times 10^{-2}\,$yr$^{-1}$
for sdB stars with normal stellar companions and $\sim 0.4\times
10^{-2}\,$yr$^{-1}$ for sdB stars with white-dwarf companions.

\subsection{The Stable Roche-Lobe Overflow (RLOF) Channel}

\begin{figure}[t]
\hspace{-1.5cm}
\mbox{\psfig{figure=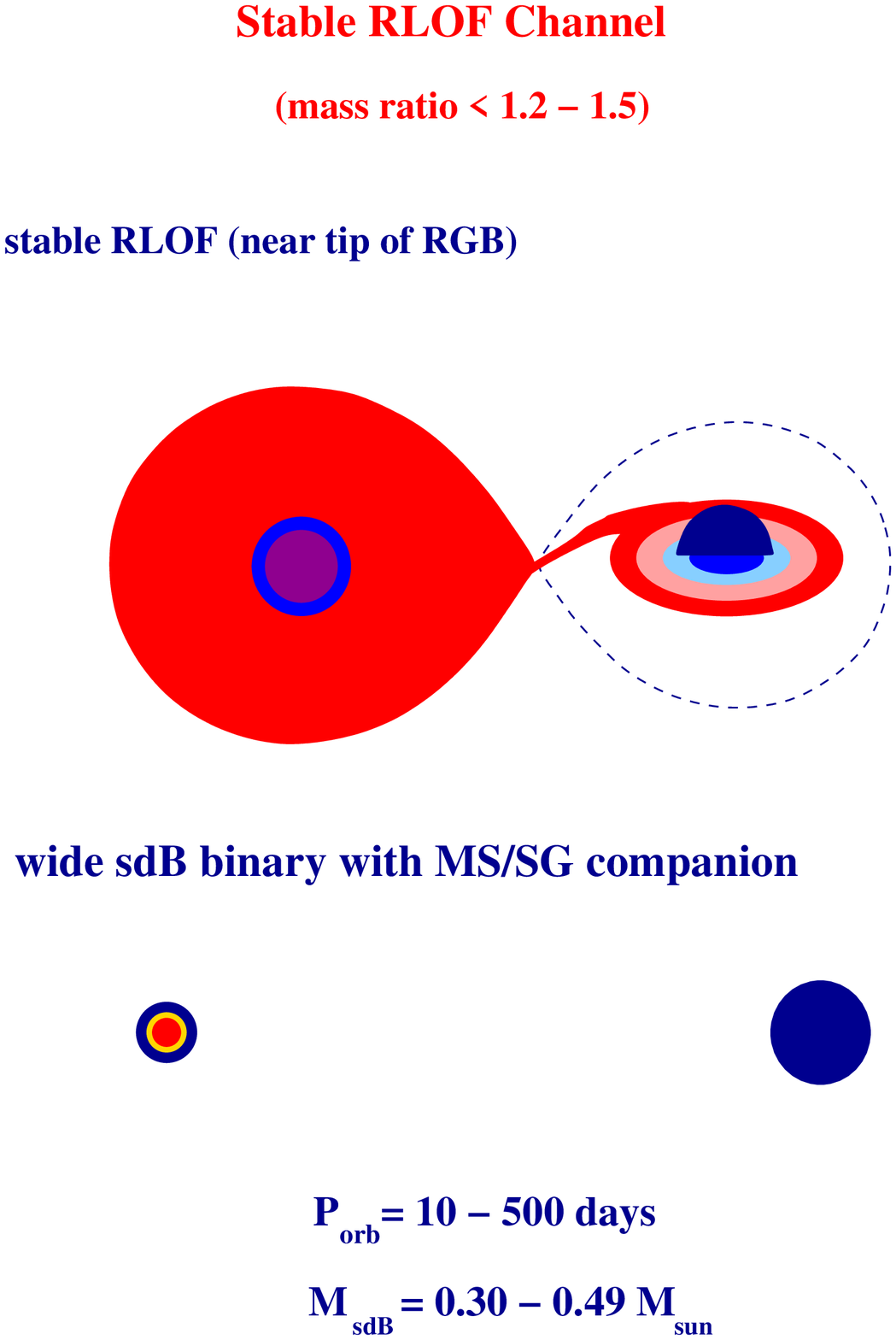,width=0.625\textwidth}
\hspace{-1cm}
\psfig{figure=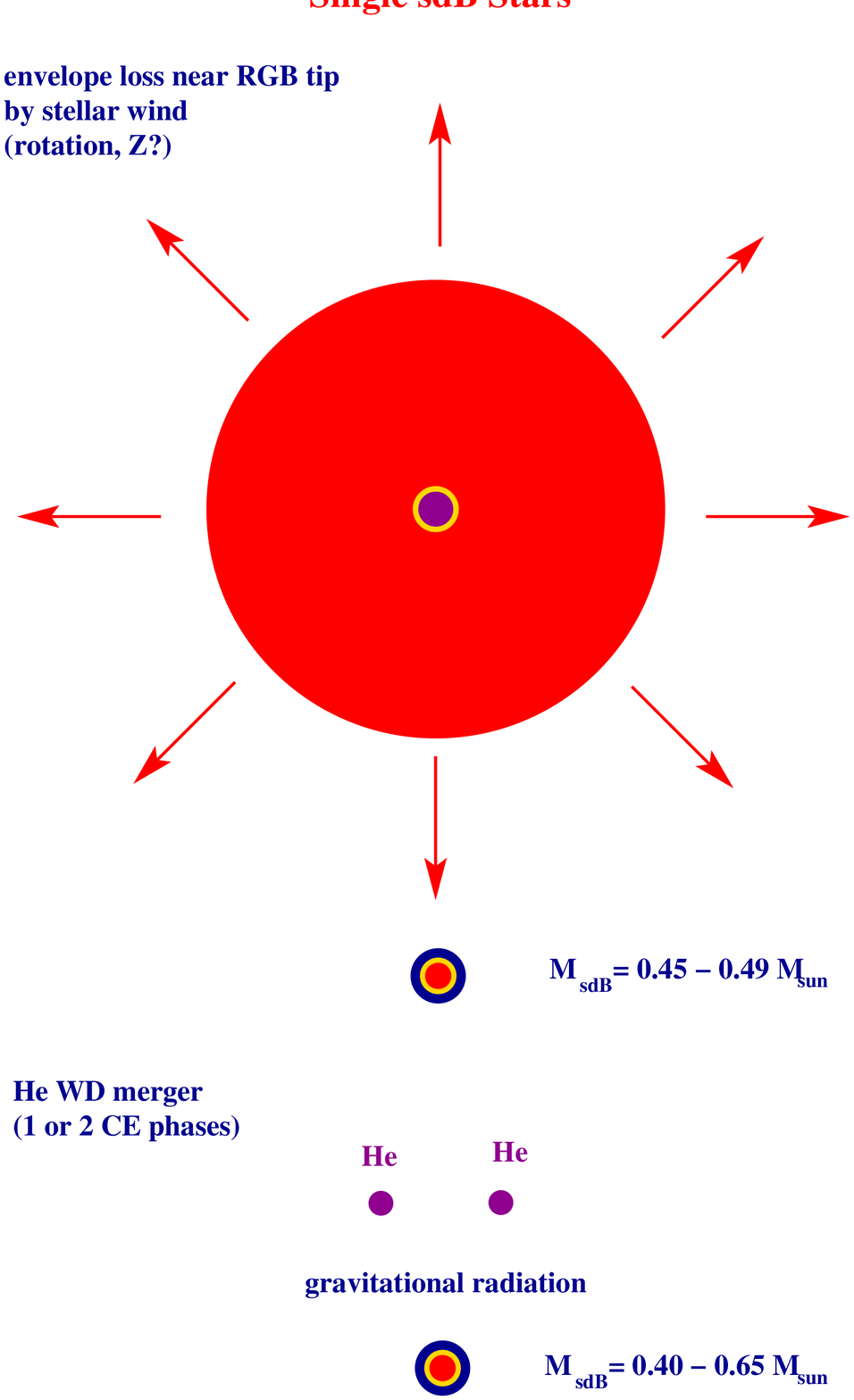,width=0.625\textwidth}}
\caption{Stable Roche-lobe channel {\em (left)} and single-star/merger 
channels {\em (right)} for the formation of sdB stars.}
\end{figure}

The main difference in the stable RLOF channel (e.g., Mengel, Norris,
\& Gross 1976) is that mass transfer (the left panel in Figure~2) is
stable and does not lead to a common-envelope and spiral-in
phase. Because of the mass transfer, the binary system tends to
widen. The typical final orbital period is 100 to 500\,d (for $M_0<
2\,M_\odot$), but can be as short as a few days for $M_0>
2\,M_\odot$. Since in the latter case, the companions are expected to
be A stars, it is likely that these systems are selected against in
most hot subdwarf surveys.

\subsection{The Merger Channel}

In the merger channel (Webbink 1984; Iben \& Tutukov 1986), the
lighter of two helium white dwarfs is dynamically disrupted when it
fills its Roche lobe. The requires that the initial orbital period of
the helium white dwarf binary is close enough ($\la 8\,$hr) that
gravitational radiation can bring the system into contact.  Most of
the mass of the disrupted star will subsequently be accreted by the
more massive white dwarf which at a certain critical mass ignites
helium and becomes a {\em single} sdB star. The distribution of sdB
stars for the merger channel is quite broad with a range from 0.4 to
0.65\,$M_\odot$. In our best-fit model, the birthrate of sdB stars in
the merger channel is $0.8- 1.6\times 10^{-2}\,$yr$^{-1}$, i.e.,  is
comparable to the CE channels.

\subsection{The Population of sdB Stars}

\begin{figure}[t]
\hspace{2cm}
\psfig{figure=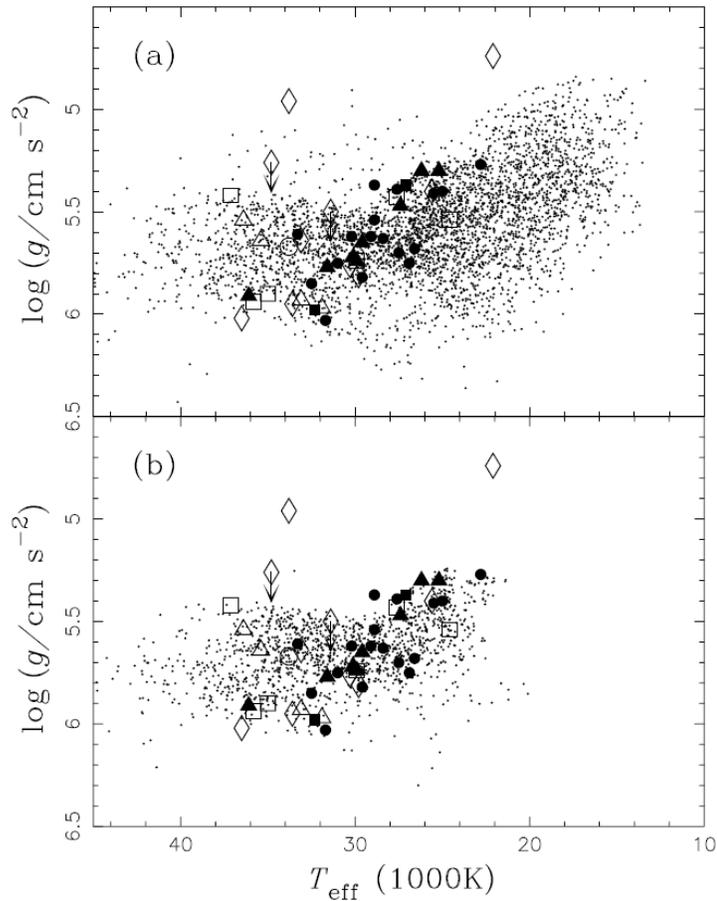,width=0.8\textwidth}
\caption{Distribution of sdB stars in the $T_{\rm eff}\,$--$\,\log g$
plane for the best-fit model of Han et al.\ (2003). Dots represent
the results of the BPS simulation. Large symbols indicate the positions
of observed various observed sdB stars. Panel (a) does not include
observational selection effects, while panel (b) does.}
\end{figure}

Figure 3 compares the results of binary population synthesis (BPS)
simulations (small dots) for the best-fit model of Han et al.\ (2003) with
the observed distribution (large symbols) in the 
$T_{\rm eff}\,$--$\,\log g$ diagram, where the
the top panel does not include observational selection effects, while
the bottom panel does. Overall the distribution of observed stars is
quite well reproduced. One of the main conclusions of these
simulations is that the birthrates of the three main evolutionary
channel (CE channel, RLOF channel and the merger channel) are of
comparable importance, consistent with recent studies (see Green 2008).

\section{The UV Excess in Elliptical Galaxies}

\begin{figure}[p]
\centering
\psfig{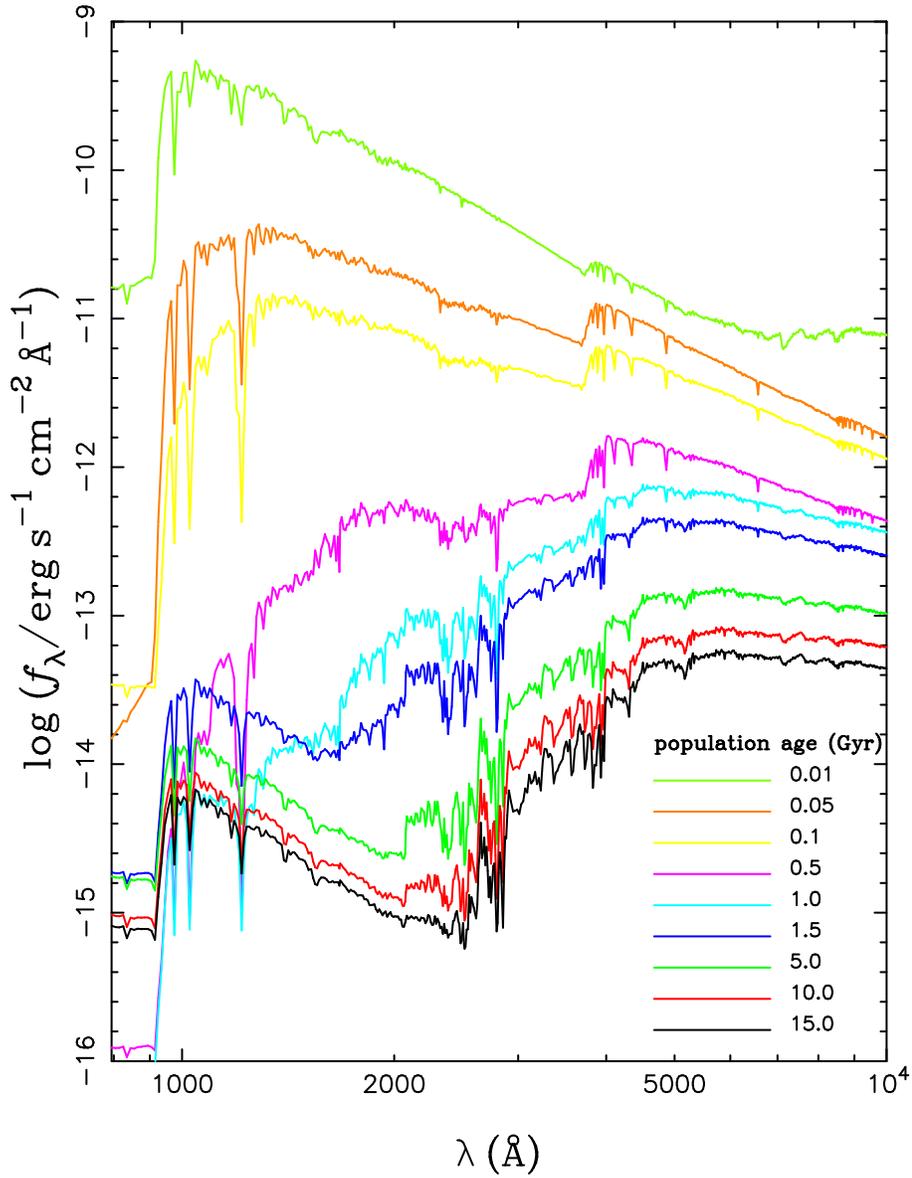}
\caption{The evolution of the far-UV spectrum with time for a single
population where all stars formed at the same time. The flux
$f_{\lambda}$ is scaled relative to the visual flux ($f_{\rm V}$).}
\label{fig1}
\end{figure}

The origin of this population of hot, blue stars in an otherwise red
population has, however, remained a major mystery (Greggio \& Renzini
1990).  Two scenarios, referred to as the high- and the
low-metallicity scenario, have been advanced. In the low-metallicity
model (Lee 1994), it is argued that these hot subdwarfs originate from
a low-metallicity population of stars which produce very blue helium
core-burning stars. This model tends to require a very large age of
the population (in fact, larger than the generally accepted age of the
Universe); it is also not clear whether the population is sufficiently
blue to account for the observed UV color. Moreover, the required low
metallicity appears to be inconsistent with the large metallicity
inferred for the majority of stars in elliptical galaxies (Terlevich
\& Forbes 2002).  In contrast, the high-metallicity model (Bressan,
Chiosi, \& Fagotto 1994; Yi, Demarque, \& Kim 1997) assumes a
relatively high metallicity -- consistent with the metallicity of
typical elliptical galaxies ($\sim 1$\,--\,3 times the solar
metallicity) -- and an associated enhancement in the helium abundance
and, most importantly, postulates an enhanced and variable mass-loss
rate on the red-giant branch, so that a fraction of stars lose most of
their hydrogen-rich envelopes before igniting helium in the core (Yi
et al.\ 1997; Dorman, O'Connell, \& Rood 1995).

Both models are quite \textit{ad hoc}: there is neither observational
evidence for a very old, low-metallicity sub-population in elliptical
galaxies, nor is there a physical explanation for the very high mass
loss required for just a small subset of stars. Furthermore, both
models require a large age for the hot component and therefore predict
that the UV excess declines rapidly with redshift. This is not
consistent with recent observations, e.g. with the Hubble Space
Telescope (HST) (Brown et al.\ 2003).  In particular, the recent
survey with the GALEX satellite (Rich et al.\ 2005) showed that the UV
excess, if anything, may increase with redshift. Indeed, the wealth of
observational data obtained with GALEX is likely to revolutionize our
understanding of elliptical galaxies.  While Burstein et al.\
(Burstein et al.\ 1988) appeared to have found a correlation between
the UV-upturn and metallicity in their sample of 24 quiescent
elliptical galaxies, which could support the high-metallicity
scenario, this correlation has not been confirmed in the much larger
GALEX sample (Rich et al.\ 2005), casting serious doubt on this
scenario.

Both models ignore the effects of binary evolution. Since we know that
in our galaxy most of the far-UV light from old stars comes from hot
subdwarfs in binaries, it is only reasonable to assume that they
will also contribute in other galaxies. Indeed, it would be implausible
to assume that the stellar populations in other galaxies were less complex
than in our own.

\section{The Model}

\begin{figure}[t]
\psfig{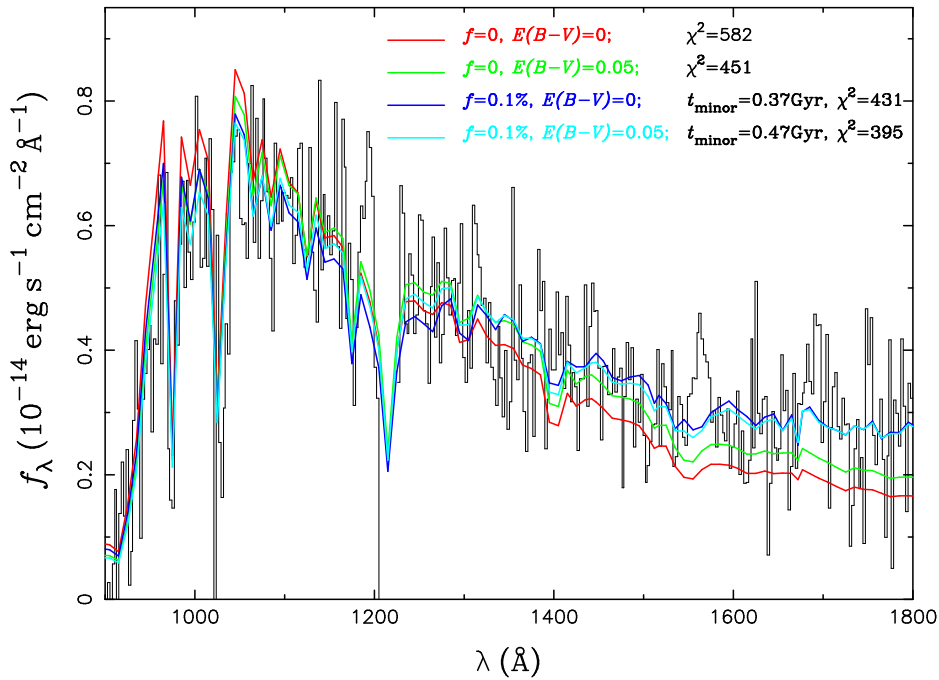}
\caption{Two-population fit to the far-UV spectrum of NGC 3379 for different 
assumptions about the fraction of the young population ($f$) and the amount
of interstellar extinction (solid curves). The grey histogram represents
the HUT observations by Brown et al.\ (1997). The theoretical models assume
the dust extinction model of Calzetti et al.\ (2000).}
\end{figure}

To quantify the importance of the effects of binary interactions on
the spectral appearance of elliptical galaxies, we have performed the
first population synthesis study of galaxies that includes binary
evolution (see also Bruzual \& Charlot 1993; Worthey 1994; Zhang, Li,
\& Han 2005).  It is based on a binary population model described
above (Han et al.\ 2002, 2003)that has been calibrated to reproduce
the short-period hot subdwarf binaries in our own Galaxy that make up
the majority of Galactic hot subdwarfs. The population synthesis model
follows the detailed time evolution of both single and binary stars,
including all binary interactions, and is capable of simulating
galaxies of arbitrary complexity, provided the star-formation history
is specified. To obtain galaxy colors and spectra, we have calculated
detailed grids of spectra for hot subdwarfs using the {\scriptsize
ATLAS9} (Kurucz 1992) stellar atmosphere code.  For the spectra and
colors of single stars with hydrogen-rich envelopes, we use the
comprehensive BaSeL library of theoretical stellar spectra
(Lejeune, Cuisinier, \& Buser 1997, 1998).

\section{Results and Discussion}

\begin{figure}[t]
\psfig{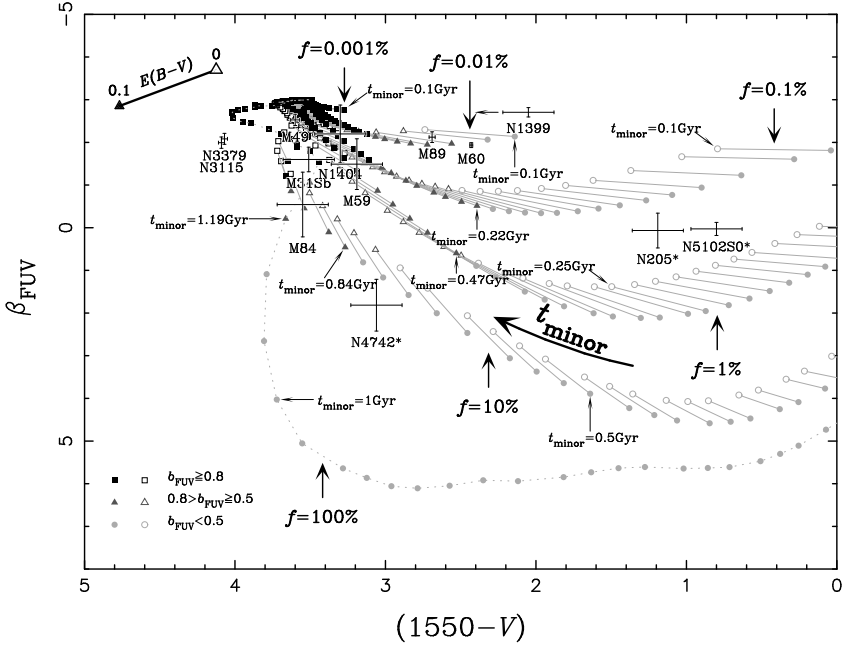}
\caption{Evolution of far-UV properties [the slope of the far-UV
spectrum, $\beta_{\rm FUV}$, versus $(1550-V)$] for a two-population
model of elliptical galaxies.  The age of the old population is
assumed to be 12\,Gyr (filled squares, filled triangles, or filled
circles) or 5\,Gyr (open squares, open triangles, or open circles).
The mass fraction of the younger population is denoted as $f$ and the
time since the formation as $t_{\rm minor}$ [plotted in steps of
$\Delta \log (t)=0.025$].  Note that the model for $f=100\%$ (the
dotted curve) shows the evolution of a simple stellar population with
age $t_{\rm minor}$.  The legend is for $b_{\rm FUV}$, which is the
fraction of the UV flux that originates from hot subdwarfs resulting
from binary interactions. The effect of internal extinction is
indicated in the top-left corner, based on the Calzetti internal
extinction model with $E(B-V)=0.1$ (Calzetti et al.\ 2000).  For
comparison, we also plot galaxies with error bars from HUT (Brown et
al.\ 1997) and IUE observations (Burstein 1988).  The galaxies with
strong signs of recent star formation are denoted with an asterisk
(NGC 205, NGC 4742, NGC 5102).  }
\label{fig2}
\end{figure}

Figure~\ref{fig1} shows our simulated evolution of the far-UV spectrum of a
galaxy in which all stars formed at the same time, where the flux has
been scaled relative to the visual flux (between 5000 and 6000\AA) to
reduce the dynamical range. At early times the far-UV flux is
dominated by the contribution from single young stars. Binary hot
subdwarfs become important after about 1.1\,Gyr, which corresponds to
the evolutionary timescale of a 2\,$M_{\odot}$ star and soon start to
dominate completely. After a few Gyr the spectrum no longer changes
appreciably.

There is increasing evidence that many elliptical galaxies had some
recent minor star-formation events (Kaviraj et al.\ 2007; Schawinski
et al.\ 2007), which also contribute to the far-UV excess.  To model
such secondary minor starbursts, we have constructed two-population
galaxy models, consisting of one old, dominant population with an
assumed age $t_{\rm old}$ and a younger population of variable age,
making up a fraction $f$ of the stellar mass of the system.  In order
to illustrate the appearance of the galaxies for different lookback
times (redshifts), we adopted two values for $t_{\rm old}$, of 12 Gyr
and 5 Gyr, respectively; these values correspond to the ages of
elliptical galaxies at a redshift of 0 and 0.9, respectively, assuming
that the initial starburst occurred at a redshift of 5 and adopting a
standard $\Lambda$CDM cosmology with $H_0=72{\rm km/s/Mpc}$,
$\Omega_{\rm M}=0.3$ and $\Omega_\Lambda=0.7$. Our spectral modelling
shows that a recent minor starburst mostly affects the slope in the
far-UV SED.  We therefore define a far-UV slope index $\beta_{\rm
FUV}$ as $f_\lambda \sim \lambda ^{\beta_{\rm FUV}}$, where
$\beta_{\rm FUV}$ is fitted between 1075\AA~ and 1750\AA. This
parameter was obtained from our theoretical models by fitting the
far-UV SEDs and was derived in a similar manner from observed far-UV
SEDs of elliptical galaxies (Burstein et al.\ 1988; Brown et al.\
1997), where we excluded the spectral region between 1175\AA~ and
1250\AA, the region containing the strong Ly$\alpha$ line.  In order
to assess the importance of binary interactions, we also defined a
binary contribution factor $b={F_{\rm b}/ F_{\rm total}}$, where
$F_{\rm b}$ is the integrated flux between 900\AA~ and 1800\AA~
radiated by hot subdwarfs produced by binary interactions, and $F_{\rm
total}$ is the total integrated flux between 900\AA ~and 1800\AA.
Figure~\ref{fig2} shows the far-UV slope as a function of UV excess, a
potentially powerful diagnostic diagram which illustrates how the UV
properties of elliptical galaxies evolve with time in a dominant old
population with a young minor sub-population.  For comparison, we also
plot observed elliptical galaxies from various sources.  Overall, the
model covers the observed range of properties reasonably well.  Note
in particular that the majority of galaxies lie in the part of the
diagram where the UV contribution from binaries is expected to
dominate (i.e. where $b> 0.5$).

The two-component models presented here are still quite simple and do
not take into account, e.g., more complex star-formation histories,
possible contributions to the UV from AGN activity, non-solar
metallicity or a range of metallicities.  Moreover, the binary
population synthesis is sensitive to uncertainties in the binary
modelling itself, in particular the mass-ratio distribution and the
condition for stable and unstable mass transfer (Han et al.\ 2003). We
have varied these parameters and found that these uncertainties do not
change the qualitative picture, but affect some of the quantitative
estimates.

Despite its simplicity, our model can successfully reproduce most of
the properties of elliptical galaxies with a UV excess: e.g., the
range of observed UV excesses, both in $(1550-V)$ and $(2000-V)$
(e.g., Deharveng, Boselli, \& Donas 2002), and their evolution with
redshift.  The model predicts that the UV excess is not a strong
function of age, and hence is not a good indicator for the age of the
dominant old population, as has been argued previously (Yi et al.\ 1999),
but is very consistent with recent GALEX findings (Rich et al.\ 2005).  We
typically find that the $(1550-V)$ color changes rapidly over the
first 1\,Gyr and only varies slowly thereafter. This also implies that
all old galaxies should show a UV excess at some level. Moreover, we
expect that the model is not very sensitive to the metallicity of the
population since metallicity does not play a significant role in the
envelope ejection process (although it may affect the properties of
the binary population in more subtle ways).

Our model is sensitive to both low levels and high levels of star
formation.  It suggests that elliptical galaxies with the largest UV
excess had some star formation activity in the relatively recent past
($\sim 1\,$Gyr ago).  AGN and supernova activity may provide
supporting evidence for this picture, since the former often appears
to be accompanied by active star formation, while supernovae, both
core collapse and thermonuclear, tend to occur mainly within
1\,--\,2\,Gyr after a starburst in the most favoured supernova models.

The modelling of the UV excess presented in this study is only a
starting point: with refinements in the spectral modelling, including
metallicity effects, and more detailed modelling of the global
evolution of the stellar population in elliptical galaxies, we suspect
that this may become a powerful new tool helping to unravel the
complex histories of elliptical galaxies that a long time ago looked
so simple and straightforward.

\acknowledgements 
This work has been supported by the Chinese National Science Foundation
under Grant Nos.\ 10433030 and 10521001 (ZH).

\end{document}